\documentclass[aps,pra,twocolumn,amsmath,amssymb,superscriptaddress, 10pt]{revtex4-2}

\usepackage{graphicx}
\usepackage{xcolor}
\definecolor{cerulean}{rgb}{0.0, 0.48, 0.65}
\definecolor{regalia}{rgb}{0.32, 0.18, 0.5}
\usepackage{accents}
\usepackage{hyperref}
\hypersetup{
    colorlinks=true,
    linkcolor=regalia,
    filecolor=regalia,      
    urlcolor=regalia,
    citecolor=regalia
}

\usepackage{braket}
\usepackage{bbold}
\usepackage{svg}
\usepackage{import}
\usepackage{bm}
\usepackage{booktabs} 
\usepackage{multirow}
\usepackage{graphicx}
\usepackage{tikz}
\usetikzlibrary{quantikz2}

\usepackage[normalem]{ulem}
\usepackage{subfigure}
\usepackage{tikz} 
 \usetikzlibrary{decorations.pathreplacing,calligraphy}

\begin{document}
\title{Fast Native Three-Qubit Gates and Fault-Tolerant Quantum Error Correction with Trapped Rydberg Ions}

\author{Katrin Bolsmann}
\email{k.bolsmann@fz-juelich.de}
\affiliation{Institute for Theoretical Nanoelectronics (PGI-2), Forschungszentrum J{\"u}lich, D-52425 J{\"u}lich, Germany}
\affiliation{Institute for Quantum Information, RWTH Aachen University, D-52056 Aachen, Germany}

\author{Thiago L. M. Guedes}
\affiliation{Institute for Theoretical Nanoelectronics (PGI-2), Forschungszentrum J{\"u}lich, D-52425 J{\"u}lich, Germany}
\affiliation{Institute for Quantum Information, RWTH Aachen University, D-52056 Aachen, Germany}

\author{Weibin Li}
\affiliation{School of Physics and Astronomy and Centre for the Mathematics
and Theoretical Physics of Quantum Non-Equilibrium Systems,
The University of Nottingham, Nottingham NG7 2RD, United Kingdom}

\author{Joseph W. P. Wilkinson}
\affiliation{Institut f{\"u}r Theoretische Physik and Center for Integrated Quantum Science and Technology, Universit{\"a}t T{\"u}bingen, D-72076 T{\"u}bingen, Germany}

\author{Igor Lesanovsky}
\affiliation{School of Physics and Astronomy and Centre for the Mathematics
and Theoretical Physics of Quantum Non-Equilibrium Systems,
The University of Nottingham, Nottingham NG7 2RD, United Kingdom}
\affiliation{Institut f{\"u}r Theoretische Physik and Center for Integrated Quantum Science and Technology, Universit{\"a}t T{\"u}bingen, D-72076 T{\"u}bingen, Germany}

\author{Markus M{\"u}ller}
\affiliation{Institute for Theoretical Nanoelectronics (PGI-2), Forschungszentrum J{\"u}lich, D-52425 J{\"u}lich, Germany}
\affiliation{Institute for Quantum Information, RWTH Aachen University, D-52056 Aachen, Germany}

\begin{abstract}
Trapped ions as one of the most promising quantum-information-processing platforms, yet conventional entangling gates mediated by collective motion remain slow and difficult to scale.  
Exciting trapped ions to high-lying electronic Rydberg states provides a promising route to overcome these limitations by enabling strong, long-range dipole--dipole interactions that support much faster multi-qubit operations.  
Here, we introduce the first scheme for implementing a native controlled-controlled-Z gate with microwave-dressed Rydberg ions by optimizing a single-pulse protocol that accounts for the finite Rydberg-state lifetime.  
The resulting gate outperforms standard decompositions into one- and two-qubit gates by achieving fidelities above $97\,\%$ under realistic conditions, with execution times of about $2\,$\textmu$\mathrm{s}$ at cryogenic temperatures.  
To explore the potential of trapped Rydberg ions for fault-tolerant quantum error correction, and to illustrate the utility of three-qubit Rydberg-ion gates in this context, we develop and analyze a proposal for fault-tolerant, measurement-free quantum error correction using the nine-qubit Bacon-Shor code.
Our simulations confirm that quantum error correction can be performed in a fully fault-tolerant manner on a linear Rydberg-ion chain despite its limited qubit connectivity. 
These results establish native multiqubit Rydberg-ion gates as a valuable resource for fast, high-fidelity quantum computing and highlight their potential for fault-tolerant quantum error correction.
\end{abstract}

\maketitle

\section{Introduction}
With exceptional controllability, long coherence times, and high connectivity, trapped ions constitute a leading platform for scalable quantum computing and simulation~\cite{Schindler2013, Bruzewicz2019, Monroe2021, Fossfeig2024, Ransford2025}.
The high fidelities achieved for one- and two-qubit gates~\cite{Campbell2010, Harty2014, Löschnauer2024, Smith2025, Zhao2025, Benhelm2008, Ballance2016} have enabled pioneering experiments in quantum error correction (QEC) and fault-tolerant (FT) quantum information processing~\cite{Chiaverini2004, Schindler2011, Linke2017, Postler2022, Berthusen2024, Postler2024, Pogorelov2025}.
However, conventional entangling-gate schemes utilizing collective vibrational modes are slow and hard to scale due to the dense motional spectrum in larger ion crystals~\cite{Bruzewicz2019}.
Faster gates driving multiple motional modes have been realized, but for larger systems closing all phase-space trajectories at the end of the driving pulse becomes increasingly challenging, leading to reduced gate fidelities~\cite{Schafer2018}.
On the other hand, Rydberg atoms interacting via van der Waals interactions enable fast, high-fidelity entangling operations~\cite{Levine2019, Henriet2020, Madjarov2020, Graham2022, Muniz2025, Tsai2025, Radnaev2025}.
Parallel entanglement of $60$ atoms at $99.5\,\%$ fidelity was realized~\cite{Evered2023}, highlighting the scalability of this platform.
Recent experiments further demonstrated the  capabilities of Rydberg atoms for implementing FT quantum-computing protocols~\cite{Rodriguez2024, Bluvstein2024, Reichardt2025, Bluvstein2025}.
However, the difficulty of providing confinement for ground and Rydberg states leads to coupling between internal and motional dynamics, which results in markedly shorter coherence times for Rydberg atoms compared with trapped-ion systems~\cite{Savard1997, Saffman2016, Robicheaux2021, Wintersperger2023}.
The limitations in both platforms can be circumvented by a novel Rydberg-ion platform that combines the precise control of trapped-ion systems with the strong, long-range interactions of Rydberg atoms~\cite{Muller2008, Mokhberi2020}.
By exciting trapped strontium $^{88}\mathrm{Sr}^+$-ions to Rydberg states, a $700\,\mathrm{ns}$ two-qubit entangling gate, much faster than conventional trapped-ion entangling operations, was recently demonstrated~\cite{Zhang2020}.
Predictions are that even faster two-qubit gates of $200\,\mathrm{ns}$ with fidelities $>99\,\%$ are achievable~\cite{Wilkinson2024}.

Given the strength of the Rydberg interactions and their reach beyond nearest neighbors, the Rydberg-ion platform naturally raises the prospect of moving beyond native two-qubit gates toward fast and robust multi-qubit operations.
Native multi-qubit operations facilitate the implementation of complex quantum algorithms by reducing the circuit depth, enabling faster execution with higher fidelity~\cite{Nemirovsky2025}.
As has been shown very recently, they are especially beneficial for QEC protocols that rely on multi-qubit stabilizer measurements~\cite{Pecorari2025, Old2025}, as well as for measurement-free QEC schemes in which they enable the efficient FT application of coherent feedback~\cite{Heußen2024, Veroni2024}.

In this work, we propose the first scheme for the implementation of a native controlled-controlled-Z (CCZ) gate with trapped Rydberg ions.
We assume that both the Rabi frequency and the detuning of the excitation laser driving the Rydberg transition are modulated in time, such that the gate is implemented by the interplay between the Rydberg interaction and the tuned laser pulses.
We maximize the fidelity through numerical optimization of the pulse parameters and analyze the gate performance for different durations.
Furthermore, we examine the impact of spontaneous decay from the Rydberg states, a primary error source in Rydberg-atom platforms, on the gate performance. 
For realistic experimental parameters, we show that the optimized native CCZ gate can be executed significantly faster than its decomposition into two-qubit entangling gates while achieving comparable fidelities.
We further investigate how varying the ratio between nearest- and next-nearest-neighbor interactions affects the gate performance.

To advance Rydberg-ion quantum processors beyond the noisy intermediate-scale quantum (NISQ) regime, where computational depth and reliability are limited by noise and decoherence, QEC protocols must eventually be incorporated~\cite{Preskill2018}.
Towards this goal, we propose a fully FT measurement-free implementation of the nine-qubit Bacon-Shor code tailored to the Rydberg-ion platform~\cite{Bacon2006}.
This measurement-free approach, which was first proposed for near-term Rydberg atom devices \cite{Veroni2024}, is particularly well suited to our system since it exploits the fast coherent operations and the native availability of nearest-neighbour three-qubit CCZ gates while avoiding slow in-sequence measurements.
We numerically demonstrate the expected quadratic scaling of the logical error rate, confirming the circuit's fault tolerance and showing that QEC protocols can be realized on a linear Rydberg-ion chain despite its limited connectivity.
Our work provides a viable route toward fast, high-fidelity native multi-qubit gates with Rydberg ions and shows that such operations can serve as useful primitives for QEC applications.

\section{Hamiltonian}
Highly excited Rydberg ions confined in a linear Paul trap must be treated as composite objects because their radius increases quadratically with the principal quantum number and, in contrast to ground-state ions, we can no longer assume that the electron experiences the same potential as the ionic core.
The Hamiltonian that describes a single trapped Rydberg ion is thus composed of three components~\cite{Muller2008, Wilkinson2024}, 
\begin{equation}\label{eqn:hamiltonian_total}
   {H = H_\mathrm{ex}+H_\mathrm{in}+H_\mathrm{co}}, 
\end{equation}
where the first two terms describe the external and internal dynamics, while the last term denotes the coupling between these two.
Similar to trapped ground-state ions, the external Hamiltonian $H_\mathrm{ex}$ describes the harmonic oscillation of the center-of-mass coordinate $\mathbf{R}$, which is predominantly determined by the heavy ionic core.
\begin{figure}[t]
    \centering
    \includegraphics[scale = 1]{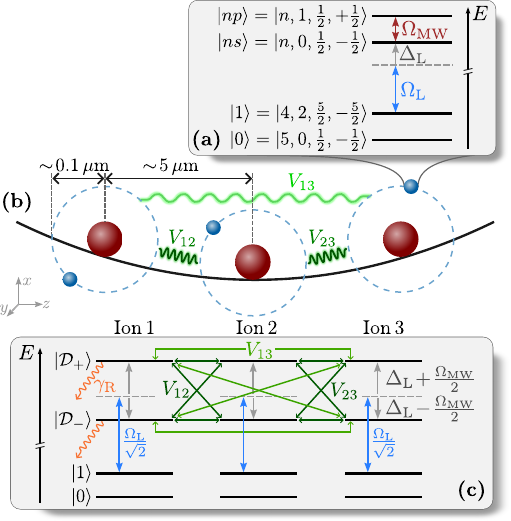}
 \caption{Electronic level structure. \textbf{(a)} Considered energy levels of a single $^{88}\mathrm{Sr}^+$-ion.
 The computational basis states are $\ket{0}$ and $\ket{1}$, with the latter coupled to the Rydberg state~$\ket{ns}$ with Rabi frequency~$\Omega_\mathrm{L}$ and detuning~$\Delta_\mathrm{L}$.
 The Rydberg states $\ket{ns}$ and $\ket{np}$ are resonantly coupled by a microwave driving field with Rabi frequency $\Omega_\mathrm{MW}$, which yields the dressed states $\ket{\mathcal{D}_\pm}$.
 \textbf{(b)} Three Rydberg ions confined in a linear Paul trap interacting via dipole--dipole interactions between nearest neighbors, $V_{12}$ and $V_{23}$, and next-nearest neighbors, $V_{13}$.
 \textbf{(c)} Complete energy-level diagram of the interacting three-ion system.
 Both dressed states $\ket{\mathcal{D}_\pm}$ are coupled to $\ket{1}$ with Rabi frequency $\frac{\Omega_\mathrm{L}}{\sqrt{2}}$ and effective laser detuning~$\frac{\Omega_\mathrm{MW}}{2}\pm\Delta_\mathrm{L}$ (cf. Hamiltonian~\eqref{eqn:Hamiltonian}).
 The decay rate $\gamma_\mathrm{R}$ accounts for the finite lifetime of the Rydberg states.
 }
  \label{fig:level_scheme}
\end{figure}

We describe the internal dynamics in terms of the relative coordinate $\mathbf{r}$ between the valence electron and the ionic core. 
The internal Hamiltonian consists of the kinetic energy of the electron, the Coulomb interaction with the ionic core $V(|\mathbf{r}|)$, and its coupling to the time-dependent trapping potential $\Phi(\mathbf{r}, t)$ and an external electric field $\mathbf{E}(t)$, yielding ${H_{\mathrm{in}} = \frac{\mathbf{p}^{2}}{2 m} + V(|{\mathbf{r}|}) - e \Phi(\mathbf{r}, t) - e \mathbf{r} \cdot \mathbf{E}(t)}$.
In absence of the trapping potential and the external field, this Hamiltonian defines the bound electronic states of the Rydberg ion ${\big[\frac{\mathbf{p}^{2}}{2 m} + V(|{\mathbf{r}|})\big]\!\ket{n,l,j,m_j} = E_{nlj}\ket{n,l,j,m_j}}$.
These states are characterized by the quantum numbers $n$, $l$, $j$, and $m_j$, denoting the principal, orbital angular momentum, total angular momentum, and magnetic degrees of freedom, respectively.
We base our discussion on the concrete level scheme shown in Fig.~\ref{fig:level_scheme}(a).
It corresponds to the electronic structure of singly ionized strontium ($^{88}\mathrm{Sr}^+$) employed in recent experiments with trapped Rydberg ions~\cite{Zhang2020}. 
Note that this choice does not limit the generality of our analysis.
The electron--trap interaction term in $H_\mathrm{in}$ leads to a quadrupole shift~$\Delta E^\mathrm{et}_{l,j,m_j}$ of the Rydberg states and additionally causes a coupling between Zeeman states with $\Delta m_j = 2$~\cite{Mokhberi2020}.
However, states with a total angular momentum $j=\frac{1}{2}$ lack permanent quadrupole moments and therefore do not experience first-order shifts~\cite{Li2014, Mokhberi2020, Wilkinson2024}.
Hence, while this effect is significant in Rydberg states with $j>\frac{1}{2}$, it is negligible for the considered Rydberg states~\cite{Higgins2017a}.
The coupling between the Zeeman levels with $\Delta m_j = 2$ induced by the trapping potential is much weaker than the fine-structure splitting.
Therefore, only couplings within a Zeeman manifold, i.e., $\Delta j = \Delta l =0$, are significant, and the considered energy levels remain unaffected~\cite{Muller2008, Higgins2017a}.
The electron interacts with the external electromagnetic field (cf. Fig.~\ref{fig:level_scheme}(a)), which comprises an excitation laser with Rabi frequency $\Omega_\mathrm{L}$ coupling the ground state~$\ket{1}$ to the Rydberg state~$\ket{ns}$ with detuning $\Delta_\mathrm{L}$, and a microwave (MW) driving field with Rabi frequency $\Omega_{\mathrm{MW}}$ resonantly coupling the two Rydberg states $\ket{ns}$ and $\ket{np}$.
Finally, in the rotating frame of the external-field and after applying the rotating-wave approximation, the single-ion electronic Hamiltonian reads
\begin{align}\label{eqn:single_ion_Hamiltonian}
    H_\mathrm{in} = &\Delta_\mathrm{L}\big(\ket{ns}\!\!\bra{ns}+\ket{np}\!\!\bra{np}\big) +\frac{\Omega_\mathrm{L}}{2}\big(\ket{ns}\!\!\bra{1}+\ket{1}\!\!\bra{ns}\big)\notag \\
    &-\frac{\Omega_\mathrm{MW}}{2}\big(\ket{ns}\!\!\bra{np}+\ket{np}\!\!\bra{ns}\big).
\end{align}

The last term in Eq.~\eqref{eqn:hamiltonian_total} describes the coupling between internal and external dynamics and gives rise to a state-dependent modification ($\sim\!n^7$) of the effective trapping frequency~\cite{Muller2008, Mallweger2025}.
Accordingly, the oscillation of the center-of-mass coordinate depends on whether the valence electron is in the ground or Rydberg state.
Throughout this work, we assume that the crystal is sufficiently cooled and the value of the principal quantum number remains moderate ($n\leq 60$).
Under these conditions, the state-dependent modification of the trapping frequency becomes less relevant and is therefore neglected~\cite{Muller2008, Zhang2020}.

We now turn from discussing a single Rydberg ion to multiple interacting Rydberg ions in a linear Paul trap as depicted in Fig.~\ref{fig:level_scheme}(b).
The Coulomb repulsion~$\sim|\mathbf{R}_{ij}|^{-1}$ between the ions $i$ and $j$, with the relative center-of-mass coordinate $\mathbf{R}_{ij} = \mathbf{R}_{i}-\mathbf{R}_{j}$, leads to the formation of a Coulomb crystal in the trap determining the equilibrium positions of the ions \cite{James1998}.
Two ions excited to Rydberg states interact via dipole--dipole interaction, 
\begin{equation}\label{eqn:dd-Hamiltonian}
   H_{\mathrm{dd},ij} = \frac{1}{4\pi\epsilon_0}\frac{\mathbf{d}_i\cdot \mathbf{d}_j - 3(\mathbf{n}_{ij}\cdot\mathbf{d}_i)(\mathbf{n}_{ij}\cdot\mathbf{d}_j)}{|\mathbf{R}_{ij}|^3},
\end{equation}
where $\mathbf{d}_i$ is the dipole operator of the $i$th ion and ${\mathbf{n}_{ij}=\frac{\mathbf{R}_{ij}}{|\mathbf{R}_{ij}|}}$.
Trapped ions in atomic eigenstates do not have permanent dipole moments and therefore the dipole--dipole interaction has no first-order effect.
The second-order effect — the van der Waals (vdW) interaction — scales as $Z^{-6}$ with the net core charge.
Compared to their neutral-atom counterparts, in trapped Rydberg ions, where $Z = 2$, this results in a strongly suppressed interaction strength ($\sim\!\mathrm{kHz}$), insufficient to enable fast entangling-gate operations~\cite{Muller2008}.
Moreover, the vdW interaction scales with $|\mathbf{R}_{ij}|^{-6}$ and in trapped-ion systems, inter-ion distances of less than a few \textmu$\mathrm{m}$ are hard to achieve due to strong Coulomb repulsion, which can trigger a transition from a linear to a zigzag configuration \cite{Fishman2008}.
However, the MW field included in the Hamiltonian~\eqref{eqn:single_ion_Hamiltonian} couples the two Rydberg states to form dressed states with permanent dipole moments rotating in the MW polarization plane, which leads to strong, long-range, first-order interactions $\sim|\mathbf{R}_{ij}|^{-3}$~\cite{Zhang2020}.
We assume the trap axis to be along the $z$-direction (cf.~Fig.~\ref{fig:level_scheme}(b)) and that the dipole moments rotate in the $xy$-plane, i.e.,~$\mathbf{n_{ij}}\cdot\mathbf{d}_i=0$.
In the basis of the dressed Rydberg states, $\ket{\mathcal{D}_{\pm}}=\frac{1}{\sqrt{2}}(\ket{ns}\mp \ket{np})$, the total Hamiltonian of interacting Rydberg ions reads~\cite{Wilkinson2024}
\begin{equation}\label{eqn:Hamiltonian}
    \begin{aligned}
        H =& \sum_{i, \pm}\Big[\Delta_\pm \! \ket{\mathcal{D}_\pm}\!\!\bra{\mathcal{D}_\pm}_{i} + \frac{\Omega_{\mathrm{L}}}{2 \sqrt{2}} \big( \ket{1}\!\!\bra{\mathcal{D}_\pm}_{i} +\text{H.c.}\big)\Big]\\
        +&\sum_{i\neq j}\!\frac{V_{ij}}{4}\big[\sigma_{z,i}^{\mathcal{D}}\sigma_{z,j}^{\mathcal{D}} + \sigma_{y,i}^{\mathcal{D}}\sigma_{y,j}^{\mathcal{D}}\big] + H_\mathrm{dec}
    \end{aligned}
\end{equation}
with detunings $\Delta_\pm = \Delta_{\mathrm{L}} \pm \frac{\Omega_{\mathrm{MW}}}{2}$, interaction strength $V_{ij} = \frac{2}{9} \frac{e^2}{4\pi\epsilon_0}\frac{|\bra{ns}r\ket{np}|^2}{|\mathbf{R}_{ij}|^3}$, and with $\sigma_{a,i}^{\mathcal{D}}$ ($a\in\{x,y,z\}$) being the Pauli matrices acting on the dressed-Rydberg subspace of the $i$th ion.
Note that the Hamiltonian~\eqref{eqn:Hamiltonian} describes global addressing, implying identical laser and microwave driving for all ions.
In the linear three-ion case we have $V_{12}=V_{23}=8V_{13} \equiv V$.
One significant source of infidelity in Rydberg-ion gates is the finite lifetime of the Rydberg states~\cite{Mokhberi2020}. 
To include this effect in our numerical analysis, we add a non-Hermitian part $H_\mathrm{dec}=-\frac{i}{2}\gamma_\mathrm{R}\sum_{i, \pm} \ket{\mathcal{D}_\pm}\!\!\bra{\mathcal{D}_\pm}_i$ to the Hamiltonian.
This provides a sufficiently accurate approximation of the population loss caused by decay, provided that the Rydberg lifetime, $\tau_\mathrm{R} = \gamma_\mathrm{R}^{-1}$, is much longer than the gate duration $\tau$ and therefore the resulting overall decay probability from the Rydberg states is small. 
The level scheme and interactions described by Hamiltonian~\eqref{eqn:Hamiltonian} are illustrated in Fig.~\ref{fig:level_scheme}(c).

\section{Implementation of three-qubit gates}
The implementation of three-qubit gates is based on the interplay between laser-driven excitation and interactions between Rydberg-excited ions, which leads to the accumulation of conditional entangling phases.
By appropriately choosing the laser pulse parameters, these phases can be controlled to realize native three-qubit gates, such as a CCZ gate.
The excitation-laser Rabi frequency and detuning are, for simplicity, modelled by sinusoidal pulses,
\begin{equation}\label{eqn:pulses}
    \Omega_\mathrm{L}(t) = \Omega_0\sin^2\Big(\frac{\pi t}{\tau}\Big),\quad
    \Delta_\mathrm{L}(t) = \delta_0 - \Delta_0\sin^2\Big(\frac{\pi t}{\tau}\Big),
\end{equation}
where the free parameters $\Omega_0$, $\delta_0$, and $\Delta_0$ remain to be optimized according to the desired gate. 
The state at the end of a pulse is given by ${\ket{\Psi(\tau)} = \sum_{a, b, c} c_{abc} e^{i \varphi_{abc}} \! \ket{abc}}$ with $(a,b,c) \in \{0, 1, \mathcal{D}_\pm\}^3$, the populations $p_{abc} = |c_{abc}|^2$, and the accumulated phases $\varphi_{abc}$.
Assuming the initial state $\ket{\Psi(t=0)} = \ket{+++}$, with $\ket{+}=\frac{1}{\sqrt{2}}(\ket{0}+\ket{1})$ the CCZ-gate fidelity $\mathcal{F} = |\!\braket{\Psi_{\mathrm{T}}|\Psi(\tau)}\!|^{2}$ quantifies the overlap with the target state $\ket{\Psi_{\mathrm{T}}} = \frac{1}{2\sqrt{2}}\sum_{i,j,k}(-1)^{ijk}\ket{ijk}$, where $(i, j, k) \in \{0, 1\}^3$ (cf. App.~\ref{app:fidelity}).
Therefore, a CCZ gate acting on $\ket{+++}$ generates GHZ-type entanglement, with any gate imperfection directly reducing the final-state fidelity.
For an ideal CCZ gate, only the coefficients for $(a,b,c) \in \{0,1\}^3$ are nonzero. 
These coefficients and associated phases fulfill the conditions
\begin{equation}\label{eqn:CCZconditions}
    c_{abc} = \frac{1}{2\sqrt{2}},\quad\varphi_{abc}^\mathrm{ent}\!\!\!\mod(2\pi) = abc\,\pi
\end{equation}
with the entangling phases $\varphi_{abc}^\mathrm{ent}=\varphi_{abc}-(a+b+c)\varphi_{100}$.
Due to the symmetry of the Hamiltonian~\eqref{eqn:Hamiltonian} under permutation of ions~1 and~3, and full permutation symmetry for initial states with a single ion in $\ket{1}$, it suffices to fulfill the conditions for $(a,b,c) \in \{(100),(101), (110), (111)\}$.
We further define the population and phase errors,
\begin{align}
    \bar{p} &= 1-\frac{1}{8}\big|\sum_{a, b, c}c_{abc}\big|^2,\\
    \bar{\varphi} &= 1 - \frac{1}{64}\big|4+2 e^{i \varphi^\text{ent}_{110}}+e^{i \varphi^\text{ent}_{101}}-e^{i \varphi^\text{ent}_{111}}\big|^2.
\end{align}
Up to second order these two error sources are independent from each other as shown in App.~\ref{app:errors}.
To implement high-fidelity CCZ gates, we minimize the infidelity $1-\mathcal{F}$ by optimizing the free parameters in Eq.~\eqref{eqn:pulses} using differential evolution \cite{Storn1997, Rocca2011}---a population-based, stochastic global optimization algorithm provided by the \texttt{scipy.optimize} library~\cite{Scipy2020}.

\begin{figure*}[t]
  \centering
  \includegraphics[scale = 1]{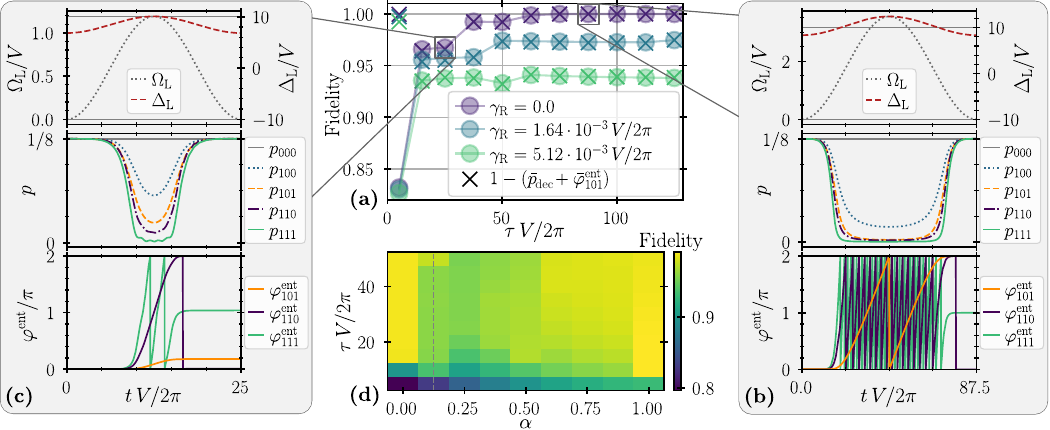}
  \caption{CCZ-gate optimization results.
  \textbf{(a)} Fidelity of the optimized CCZ-gate protocols~\eqref{eqn:pulses} vs. the dimensionless gate time~$\tau V$ for different decay rates $\gamma_\mathrm{R}$.
  Crosses ($\times$) represent approximate fidelities considering only next-nearest neighbor entangling-phase errors $\bar{\varphi}_{101}^\mathrm{ent}$ and decay-induced population errors $\bar{p}_\mathrm{dec}$.
  \textbf{(b)} Gate dynamics with $\mathcal{F}\approx 1$ and $\gamma_\mathrm{R}=0$.
  Top: Modulation of the Rabi frequency $\Omega_\mathrm{L}$ and the detuning $\Delta_\mathrm{L}$ of the excitation laser. 
  The solid gray lines at $\Delta_\mathrm{L}=\mp\frac{\Omega_\mathrm{MW}}{2}$ indicate resonance with the dressed Rydberg states $\ket{\mathcal{D}_\pm}$.
  Center: Population dynamics of the computational basis states for the initial state~$\ket{+++}$.
  Bottom: Accumulated entangling phases $\varphi_{abc}^\mathrm{ent}$.
  \textbf{(c)} Same as in (b) for a gate with reduced fidelity $\mathcal{F}\approx 96.75\,\%$ due to shorter gate time $\tau V$.
  \textbf{(d)} Gate fidelity for various gate times vs.~the interaction ratio $\alpha = \frac{V_{13}}{V}$ for $\gamma_\mathrm{R}=1.64\cdot10^{-3}\,V/2\pi$. 
  The dashed gray line indicates the linear Coulomb crystal configuration ($V_{13} = \frac{V}{8}$).
  The optimized parameters, gate fidelities, population and phase errors of the gates can be found in App.~\ref{app:parameter}.
  }
  \label{fig:gates}
\end{figure*}

Figure~\ref{fig:gates}(a) depicts the CCZ-gate fidelity after optimizing the pulses~\eqref{eqn:pulses} for various gate durations $\tau$ and decay rates $\gamma_\mathrm{R}$.
We consider three scenarios: one optimal case without Rydberg decay, i.e, $\gamma_\mathrm{R}= 0$, and two cases with $\gamma_\mathrm{R}\neq 0$.
Assuming $V=2\pi\times 25\,\mathrm{MHz}$, which can be achieved for a principal quantum number of $n=46$ ($n=60$) and an inter-ion distance of $|\mathbf{R}_{ij}|=2.3\,\text{\textmu}\mathrm{m}$ ($|\mathbf{R}_{ij}|=3.3\,\text{\textmu}\mathrm{m}$), we have $\gamma_\mathrm{R} = 0.128\,\mathrm{MHz}$ for the green and $\gamma_\mathrm{R} = 0.041\,\mathrm{MHz}$ for the blue data points in Fig.~\ref{fig:gates}(a).
The first decay rate corresponds to a room-temperature setup ($T = 300\,\mathrm{K}$) with $n=46$ as in Ref.~\cite{Zhang2020}, while the other describes an improved cryogenic scenario ($T = 0\,\mathrm{K}$) with ${n=60}$.
The two main sources of infidelity are the phase error dominated by $\varphi^\mathrm{ent}_{101}$, which enters as $\bar{\varphi}^\mathrm{ent}_{101} = 1 - \frac{1}{64}\big|7+e^{i \varphi^\text{ent}_{101}}\big|^2$, and the population error $\bar{p}_\mathrm{dec}$ caused by Rydberg decay.
This can be estimated by $\bar{p}_\mathrm{dec} \approx 1- |1 - \frac{\gamma_\mathrm{R}}{2}\int_0^\tau\!\! \text{d} t\sum_{i, \pm}p_{\pm,i}|^2,$ where $p_{\pm, i}$ denotes the time-dependent $\ket{\mathcal{D}_\pm}$-state population of the $i$th ion.

\begin{figure}[t]
  \centering
  \includegraphics[scale = 1]{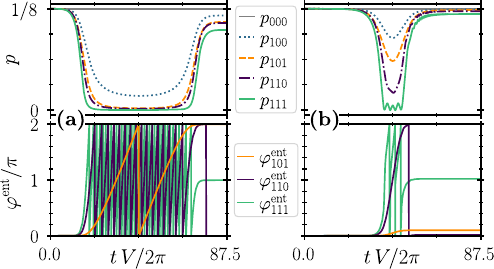}
  \caption{Comparison of CCZ-gate optimization with and without decay.
  Depicted are the population dynamics for initial state $\ket{+++}$ (top) and accumulated entangling phases~$\varphi_{abc}^\mathrm{ent}$~(bottom) both simulated with a decay rate ${\gamma_\mathrm{R}=1.64\cdot10^{-3}\,V/2\pi}$.
  \textbf{(a)}~Gate optimization without accounting for decay results in significant population loss and a reduced fidelity of $\mathcal{F}=89.65\,\%$. 
  \textbf{(b)}~The same gate optimized while accounting for decay. The population loss is decreased by reducing the time spent in the Rydberg states, yielding a substantially improved fidelity of $\mathcal{F}=97.29\,\%$.
  The optimized parameters, gate fidelities, population loss and phase errors of the gates can be found in App.~\ref{app:parameter}.
  }
  \label{fig:decay}
\end{figure}

In the absence of decay, we observe two characteristic dynamical behaviours in the CCZ-gate implementation.
For longer gate times, e.g., $\tau V/2\pi =87.5$, the ions can spend a sufficient amount of time in the Rydberg states to accumulate the desired phases obeying the conditions~\eqref{eqn:CCZconditions}, cf. Fig.~\ref{fig:gates}(b).
In this case, the detuning is modulated such that the excitation laser is on resonance with $\ket{\mathcal{D}_-}$ twice during the gate (upper plot), first bringing most of the population to the Rydberg states and then transferring it back.
In between, the laser gets far off-resonance and we observe a population plateau (center plot).
During that time, the ions interact through the dipole--dipole interaction and accumulate the desired entangling phases (lower plot) resulting in a low infidelity $\sim 10^{-7}$.
Since $V_{13}=\frac{V}{8}$, the phase $\varphi^\mathrm{ent}_{101}$ accumulates the slowest and while the other entangling phases perform several rotations in the interval $[0, 2\pi]$, it only rotates twice until reaching the desired value.
When decreasing $\tau V$, the phase $\varphi^\mathrm{ent}_{101}$ cannot perform a full rotation, fundamentally limiting the gate fidelity.
Therefore, for shorter gates times, e.g., $\tau V/2\pi =25$, we observe a different characteristic dynamical behaviour as shown in Fig.~\ref{fig:gates}(c).
Here, the time that the ions spend in the Rydberg states is just sufficient to generate the required phases $\varphi^\mathrm{ent}_{110}$ and $\varphi^\mathrm{ent}_{111}$, while $\varphi^\mathrm{ent}_{101}$ is kept as small as possible.
As a result, the residual phase error decreases the fidelity to $96.75\,\%$.

It is also advantageous to minimize the time spent in the Rydberg states when considering their finite lifetime. 
We find that, when decay is included, the optimization yields dynamics similar to those observed for short gate times.
This is illustrated in Fig.~\ref{fig:decay}, which compares the performance under consideration of a decay rate $\gamma_\mathrm{R}=1.64\cdot10^{-3}\,V/2\pi$ of the gate from Fig.~\ref{fig:gates}(b), optimized without accounting for decay, to a gate optimized with decay.
The reduced time spent in the Rydberg states leads to a population loss of only $\bar{p}\approx0.02$ for the decay-optimized gate, compared to $\bar{p}\approx0.1$ in the other case, with an improved fidelity of $97.29\,\%$ instead of $89.65\,\%$. 
This comes at the cost of a slightly increased phase error of $\bar{\varphi}\approx 0.01$.
All optimized parameters, gate fidelities, population and phase errors of the discussed gates can be found in App.~\ref{app:parameter}.

Our analysis shows that $\varphi^\mathrm{ent}_{101}$ fundamentally influences the gate fidelity and nature of the implemented gate.
Without decay and assuming that the phase error is determined exclusively by the phase $\varphi_{101}^\mathrm{ent}$, the implemented gate $U$ can be written as a coherent combination of a CCZ and a next-nearest-neighbor CZ gate,
\begin{equation}
U = e^{i\varphi_{101}^\mathrm{ent}/2}\Big[\cos\Big(\frac{\varphi_{101}^\mathrm{ent}}{2}\Big)\mathrm{CCZ} - i \sin\Big(\frac{\varphi_{101}^\mathrm{ent}}{2}\Big)\mathrm{C}_1\mathrm{Z}_3 \Big].
\end{equation}
Here, $\mathrm{C}_1\mathrm{Z}_3$ denotes a CZ operation between the two outer ions of the chain.
This motivates the question whether a native $\mathrm{C}_1\mathrm{Z}_3$ gate might emerge more naturally in our setup than the CCZ gate.
Our numerical investigations (cf. App.~\ref{app:C1Z3} and Tab.~\ref{tab:performance}) indicate that high-fidelity $\mathrm{C}_1\mathrm{Z}_3$ gates are more demanding than CCZ-gate implementations, reaching lower fidelity for longer gate times.

Furthermore, we investigate the gate fidelity as a function of the interaction ratio $\alpha = \frac{V_{13}}{V}$ to gain fundamental insight into the underlying gate mechanism.
This ratio governs the accumulation rate of $\varphi^\mathrm{ent}_{101}$ and may become relevant in higher-dimensional ion configurations, which are beyond the scope of the present model.
Figure~\ref{fig:gates}(d) shows the resulting gate-fidelity dependence on the gate duration $\tau$ and the interaction ratio $\alpha$ for a decay rate $\gamma_\mathrm{R}=1.64\cdot10^{-3}\,V/2\pi$.
For very short gate times, the fidelity remains low for all values of $\alpha$, as the required entangling phases cannot be accumulated.
We observe high fidelities $\mathcal{F}>98\,\%$ at the extremes $\alpha\in \{0,1\}$. 
For $\alpha = 0$, we have $\varphi^\mathrm{ent}_{101} = 0$ during the entire pulse automatically satisfying the corresponding phase condition~\eqref{eqn:CCZconditions}.
At $\alpha = 1$, the equal interaction strengths lead to $\varphi^\mathrm{ent}_{101} = \varphi^\mathrm{ent}_{110}$, ensuring that the phase condition for $\varphi^\mathrm{ent}_{101}$ is fulfilled whenever the condition for $\varphi^\mathrm{ent}_{110}$ is satisfied.
Additionally, for $\alpha = 1$, the faster accumulation of $\varphi^\mathrm{ent}_{111}$ enables higher fidelities at shorter gate durations.

For benchmarking, we assume ${V = 2\pi\times 25\,\mathrm{MHz}}$, which can be achieved for principal quantum numbers $n\leq 60$ and typical inter-ion distances (see discussion above).  
Then the two decay rates used in Fig.~\ref{fig:gates}, $\gamma_\mathrm{R} \in \{0.128, 0.041\}\,\mathrm{MHz}$, correspond to realistic room-temperature ($T=300\,\mathrm{K}$) and cryogenic ($T=0\,\mathrm{K}$) conditions, respectively.  
A textbook decomposition of a CCZ gate involves six CZ gates, four between nearest neighbours and two between next-nearest neighbours, together with several single-qubit operations~\cite{Shende2009} (see App.~\ref{app:decom}).  
The fidelities of the next-nearest-neighbour $\mathrm{C}_1\mathrm{Z}_3$ gate, which can be realized either with local addressing (targeting only ions~1 and~3) or with global addressing (driving all ions simultaneously), are considerably lower than those of the nearest-neighbour CZ gate, as shown in Tab.~\ref{tab:performance}.  
Consequently, under room-temperature (cryogenic) conditions, six (five) nearest-neighbour CZ gates combined still outperform the fidelity of a single next-nearest-neighbour CZ gate.  
Hence, we also consider a decomposition based solely on nearest-neighbour interactions, which consists of eight nearest-neighbour CZ gates and single-qubit rotations, for benchmarking, as detailed in App.~\ref{app:decom}.  
As current experimental performances in trapped ground-state ions demonstrate single-qubit gate error rates at least one order of magnitude below those of entangling Rydberg-ion gates ~\cite{Campbell2010, Harty2014, Smith2025, Ransford2025}, we assume single-qubit gates to be error-free and take an execution time of approximately $1\,\text{\textmu}\mathrm{s}$~\cite{Strohm2024}. 
The resulting fidelities and execution times of the two CCZ decompositions are summarized in Tab.~\ref{tab:performance}.  
In terms of fidelity, the native CCZ gate performs comparably to the decomposed ones under both room-temperature and cryogenic conditions.  
However, its execution time is reduced by more than $8\,\text{\textmu}\mathrm{s}$ compared to the decomposed variants.
Moreover, the native CCZ gate can be implemented using global laser addressing, which makes it experimentally less demanding than schemes that rely on individual ion control.  
These results highlight both the performance and practical advantages of the native CCZ-gate implementation. 

\begin{table}[t]
\begin{tabular}{@{}l@{\hspace{0.2cm}}l@{\hspace{0.3cm}}ccc@{\hspace{0.3cm}}cc}
                                    & & \multicolumn{2}{c}{$\bm{T = 300}\,\mathbf{K}$}  && \multicolumn{2}{c}{$\bm{T = 0}\,\mathbf{K}$}      \\[0.1cm]
                                    & \textbf{gate}                                & $\tau[\text{\textmu}\mathrm{s}]$ & $\mathcal{F}[\%]$ && $\tau[\text{\textmu}\mathrm{s}]$ & $\mathcal{F}[\%]$ \\ \midrule
\multirow{4}{*}{\rotatebox[origin=c]{0}{\textbf{local}}} 
& $\mathrm{C}_1\mathrm{Z}_2$          & $0.2$                  & $99.22$  &   & $0.2$                  & $99.75$          \\
& $\mathrm{C}_1\mathrm{Z}_3$          & $0.7$                  & $95.54$  &   & $0.8$                  & $98.61$          \\
& CCZ (TB decomp.)                      & $9.2$                  & $88.46$  &   & $9.4$                 & $96.27$          \\
& CCZ (NN decomp.)                       & $10.6$                 & $93.93$  &   & $10.6$                 & $98.02$          \\\midrule
\multirow{2}{*}{\rotatebox[origin=c]{0}{\textbf{global}}} 
& $\mathrm{C}_1\mathrm{Z}_3$          & $2.0$                  & $90.31$  &   & $2.0$                  & $96.53$          \\ 
& CCZ                                 & $1.5$                  & $93.85$  &   & $2.0$                  & $97.42$          \\\bottomrule
\end{tabular}
\caption{Gate fidelities and durations for optimized quantum gates at different temperatures, i.e., different Rydberg-state lifetimes.  
We distinguish between gates implemented with local and global ion addressing.  
Local implementations of the CCZ gate correspond to decompositions into single- and two-qubit gates, either following the textbook construction or using only nearest-neighbour two-qubit gates (cf.~App.~\ref{app:decom}).  
The best fidelities were selected for $\tau \leq 1\,\text{\textmu}\mathrm{s}$ (local) and $\tau \leq 2\,\text{\textmu}\mathrm{s}$ (global).  
The native CCZ gate clearly outperforms the decomposed realizations in terms of execution time while having similar fidelities.}
\label{tab:performance}
\end{table}

\section{Implementation of a measurement-free Bacon-Shor code with Rydberg ions}
Quantum error correction is required for Rydberg-ion quantum processors to overcome the limitations of the NISQ regime and enable scalable operation.
The Rydberg-ion platform offers fast coherent gate operations, allowing for potentially short QEC cycle times, while in-sequence measurements remain comparatively slow~\cite{Higgins2018, Zhang2020}.
This naturally motivates the implementation of measurement-free QEC schemes that process error information coherently.
A fully FT measurement-free implementation of the nine-qubit Bacon-Shor code has recently been proposed for implementation on near-term Rydberg-atom hardware~\cite{Veroni2024}.
This approach is well suited to the Rydberg-ion platform despite its limited connectivity, as it relies on three-qubit CCZ gates which, through the proposal in this work, are natively available.
In the following, we present a fully FT QEC scheme for Rydberg ions by mapping the measurement-free nine-qubit Bacon-Shor code onto the native Rydberg-ion gate set, comprising single-qubit gates, nearest-neighbour and next-nearest-neighbour CZ gates, and nearest-neighbour three-qubit CCZ gates along a linear ion string (cf. Tab.~\ref{tab:performance}).

In general, a protocol for a distance-3 QEC code, that is able to correct at least one arbitrary error on any of its constituent physical qubits, is considered FT if the quantum circuits realizing the QEC protocols are designed such that a single physical fault cannot propagate into an uncorrectable error on the logical qubit.
As a consequence, logical failures can only arise from pairs of independent faults, implying that the logical error rate must scale at least quadratically in the physical error probability~\cite{Gottesmann2024}.
In measurement-free implementations, stabilizer information is coherently mapped onto ancillary qubits and the correction is applied via coherent quantum feedback, thereby avoiding the in-sequence measurements typically used in QEC implementations.
By eliminating the slow measurements, which often dominate the overall timescale, measurement-free schemes are particularly well suited to fast Rydberg-based architectures.
The coherent correction can be efficiently realized using multi-controlled gates, making the CCZ gate proposed in this work especially attractive for implementing such protocols.

\begin{figure*}[t]
  \centering
  \includegraphics[width=\linewidth]{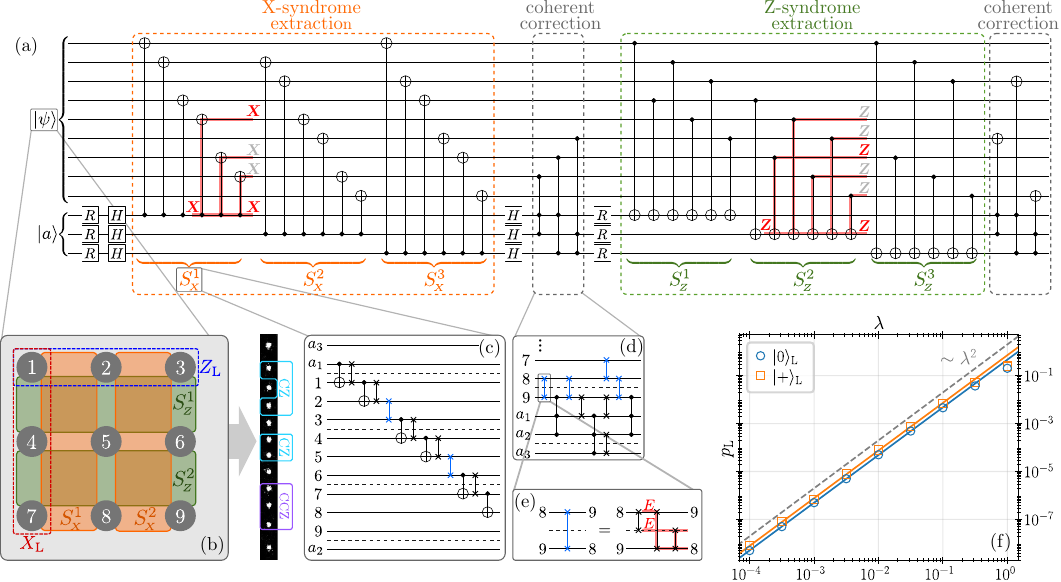}
  \caption{Measurement-free FT QEC protocol for the nine-qubit Bacon-Shor code and its implementation with the available Rydberg-ion gate set.
    \textbf{(a)}~Measurement-free FT protocol of the nine-qubit Bacon-Shor code without connectivity restrictions~\cite{Veroni2024}. Fault tolerance is ensured by extracting redundant syndrome information and by an appropriate ordering of the stabilizer readouts, as discussed in the main text. The illustrated error-propagation examples show that hook errors map only onto gauge operators, possibly accompanied by at most one correctable weight-one data error.
    \textbf{(b)}~Nine-qubit Bacon-Shor code on a $3\times3$ lattice including the stabilizers $S_X^{1,2}$ and $S_Z^{1,2}$ together with the logical operators $X_\mathrm{L}$ and $Z_\mathrm{L}$.
    \textbf{(c)}~Realization of the $S_X^{1}$-readout with the available Rydberg-ion gate set using additional SWAP operations. Dashed lines indicate SWAP ancillas required for FT SWAPs (blue).
    \textbf{(d)}~Realization of the coherent correction with the available Rydberg-ion gate set using additional SWAP operations and native CCZ gates.
    \textbf{(e)} Fault-tolerant SWAP operation in which the exchange between two qubits is mediated via an ancilla (dashed line), preventing correlated errors on the two data qubits (cf.\ red error propagation).
    \textbf{(f)}~Logical error rate $p_\mathrm{L}$ as a function of the noise-scaling parameter $\lambda$ for logical states initialized in $\ket{0}_\mathrm{L}$ and $\ket{+}_\mathrm{L}$. The dashed line indicates a $\lambda^2$ dependence, illustrating the expected quadratic scaling of $p_\mathrm{L}$ characteristic of an FT circuit. A power-law fit of the data, excluding the two largest values of $\lambda$, yields $p_\mathrm{L}(\lambda) = C \lambda^\alpha$ with $\alpha \approx 1.99$ for both logical states.}
  \label{fig:FT_BSC}
\end{figure*}

The distance-3 Bacon-Shor code \cite{Bacon2006} encodes one logical qubit into nine physical qubits which can be thought of as being arranged on a ${3\times3}$~lattice, as illustrated in Fig.~\ref{fig:FT_BSC}(b).  
Its code space is defined as the joint $+1$ eigenspace of four stabilizers,
\begin{equation}
\begin{aligned}\label{eqn:stabilizer}
S_X^1 &= X_1 X_2 X_4 X_5 X_7 X_8, \\
S_X^2 &= X_2 X_3 X_5 X_6 X_8 X_9, \\
S_Z^1 &= Z_1 Z_2 Z_3 Z_4 Z_5 Z_6, \\
S_Z^2 &= Z_4 Z_5 Z_6 Z_7 Z_8 Z_9,
\end{aligned}
\end{equation}
which act on adjacent rows and columns of the lattice, respectively. 
Therefore, the logical states can be expressed either as row- or column-encoded GHZ-like states,
\begin{align}
|0/1\rangle_\mathrm{L} &= \tfrac{1}{\sqrt{2}}\big(|000\rangle \pm |111\rangle\big)^{\otimes3\,\text{rows}},\label{eqn:log_0}\\
|\pm\rangle_\mathrm{L} &= \tfrac{1}{\sqrt{2}}\big(|+++\rangle \pm |---\rangle\big)^{\otimes3\,\text{cols}},\label{eqn:log_+}
\end{align}
which can be encoded by a unitary FT encoding circuit (cf. Ref.~\cite{Egan2021} and App.~\ref{app:BS-circuit}).
The corresponding logical operators can be chosen as $X_\mathrm{L} = X_1 X_4 X_7$ and $Z_\mathrm{L} = Z_1 Z_2 Z_3$, which commute with all stabilizers and therefore preserve the code space.
The Bacon-Shor code is defined only up to a group of gauge operators, 
\begin{align}
X_j X_{j+1}, \qquad j \in \{1,2,4,5,7,8\},\\
Z_i Z_{i+3}, \qquad i \in \{1,2,3,4,5,6\},
\end{align}
generated by pairwise products of Pauli $X$-($Z$-)operators acting on the same row (column).  
Applying any of these gauge operators does not affect the logical information since they commute with both the logical operators and stabilizers.
Furthermore, the Bacon-Shor code supports a transversal logical Hadamard gate up to qubit permutations $(2\leftrightarrow4,\,3\leftrightarrow7,\,6\leftrightarrow8)$, corresponding to a reflection of the lattice along its diagonal.

An FT realization of a measurement-free Bacon-Shor QEC cycle correcting both bit and phase flip errors, not yet subject to qubit connectivity constraints, is shown in Fig.~\ref{fig:FT_BSC}(a).
Here, fault tolerance is ensured by introducing redundancy in the extracted syndrome information and by an appropriate ordering of the stabilizer readout sequence \cite{Crow2016, Veroni2024}.
Beyond the minimal generating set~\eqref{eqn:stabilizer}, two redundant stabilizers,
\begin{equation}
\begin{aligned}
S_X^3 &= S_X^1 S_X^2 = X_1 X_3 X_4 X_6 X_7 X_9, \\
S_Z^3 &= S_Z^1 S_Z^2 = Z_1 Z_2 Z_3 Z_7 Z_8 Z_9,
\end{aligned}
\end{equation}
are extracted.
The redundant syndromes provide cross-validation between stabilizer measurements and ensure that errors on ancillas do not trigger false coherent corrections.
Moreover, fault tolerance requires a specific sequential ordering of the stabilizer readouts, in which each stabilizer is extracted as triplets of gauge operators to confine error propagation to correctable subsets.
For instance, the stabilizer $S_Z^2$ must be read out in the order $Z_4 Z_7 Z_5 Z_8 Z_6 Z_9$.
As illustrated in Fig.~\ref{fig:FT_BSC}(a), this ordering confines hook errors, i.e., errors propagating from an ancilla to data qubits, to gauge operators, possibly accompanied by a single-qubit error, but never to an uncorrectable multi-qubit error.

To embed the discussed FT circuit into a one-dimensional Rydberg-ion string with only nearest- and next-nearest-neighbour connectivity, additional SWAP operations are inserted to realize all the desired two- and three-qubit operations.
Figure~\ref{fig:FT_BSC}(c) and (d) show this construction exemplarily for the $S_X^1$-readout and a coherent correction step.
One additional challenge here is that, in general, SWAP gates are not FT, as noise during their operation can induce correlated errors on both participating qubits, leading to uncorrectable errors.
We therefore introduce FT SWAP operations (cf. Fig.~\ref{fig:FT_BSC}(e)), in which the exchange of two qubits is mediated through an auxiliary qubit~\cite{DeJong2019}.
This circuit is constructed such that any correlated error arising during the SWAP operation propagates to at most one of the two data qubits, and never to both simultaneously.
To implement these FT SWAPs on a linear ion string, we insert dedicated SWAP ancillas after every third qubit, enabling all required permutations to be performed fault-tolerantly.
Remarkably, as shown in Fig.~\ref{fig:FT_BSC}(c) and (d), we find that not all SWAP operations need to be implemented fault-tolerantly in order to maintain overall fault tolerance of the QEC cycle.
Fault-tolerant SWAPs are always required when exchanging two data qubits, since otherwise an uncorrectable weight-two error on both data qubits can appear.
SWAPs between ancillas, in contrast, can typically be performed non-fault-tolerantly, since correlated ancilla errors either result in a false correction corresponding to a correctable weight-one data error or manifest as hook errors that, as discussed above, remain within the correctable set of errors.
For SWAPs between ancillas and data qubits, FT execution is required only if a correlated error could change the syndrome in a way that causes an incorrect correction.
The full implementation of the optimized FT measurement-free Bacon-Shor circuit adapted to the restricted Rydberg-ion connectivity is shown in App.~\ref{app:BS-circuit}.
Here, FT SWAPs are used only where strictly required to preserve fault tolerance.
Compared to a naive implementation employing FT SWAPs throughout, the total number of physical CNOT gates is reduced by approximately $25\,\%$.
However, it is not guaranteed that this circuit is optimal in terms of gate count and circuit depth. 
Given the high complexity and large optimization space of this optimization problem, a more systematic, potentially automatized, improvement of the QEC quantum circuit is left for future work.

To analyze its performance, we perform a density matrix simulation of the QEC circuit under depolarizing noise using the Python package \textsc{Cirq}~\cite{Cirq}.
We consider a single round of error correction (i.e.~a single QEC cycle) and assume, for simplicity, ideal initial logical state encoding. 
The depolarizing channel is defined as  
\begin{equation}
    \mathcal{E}(\rho) = (1-p)\rho + \frac{p}{4^n - 1}\sum_i P_i \rho P_i,
\end{equation}
where $P_i$ are the $4^n - 1$ nontrivial Pauli operators. 
Accordingly, after an $n$-qubit gate with fidelity $F$, a Pauli error appears with probability $p = \frac{2^n + 1}{2^n}(1 - F)$~\cite{Nielsen2002}.  
Single-qubit gates are assumed to be error-free, while the depolarizing channel with error rates $\mathbf{p}_\mathrm{phys} = (p_2^\mathrm{nn}, p_2^\mathrm{nnn}, p_3)$, calculated from the gate fidelities obtained under cryogenic conditions as listed in Tab.~\ref{tab:performance}, is applied to all nearest- and next-nearest-neighbour two-qubit and three-qubit gates. 
To study the scaling of the logical error rate, we rescale $\mathbf{p}_\mathrm{phys} \rightarrow \lambda \mathbf{p}_\mathrm{phys}$ with $\lambda \in (0,1]$, thereby obtaining a simple one-parameter model for anticipated future reductions in physical gate error rates.
We then plot the resulting logical error rate $p_\mathrm{L}$ for the initial states $\ket{0}_\mathrm{L}$ and $\ket{+}_\mathrm{L} =\frac{1}{\sqrt{2}}(\ket{0}_\mathrm{L}+\ket{1}_\mathrm{L}) $ as a function of $\lambda$ in Fig.~\ref{fig:FT_BSC}(f).
As expected for a FT circuit, the logical error rate scales quadratically with the scaling parameter $\lambda$, confirming, as expected, that all single faults remain correctable within the implemented QEC scheme.

\section{Conclusion}
We have theoretically demonstrated the feasibility of implementing native CCZ gates using trapped Rydberg ions within experimentally reasonable parameter ranges.
Taking into consideration the finite lifetime of Rydberg states at both room-temperature and cryogenic conditions, the proposed native CCZ gate achieves comparable fidelities while clearly outperforming its decomposed alternatives in terms of execution time.
An additional observation arising from the analysis of the dominant error sources is that the implemented gates are a coherent combination of CCZ and $\mathrm{C}_1\mathrm{Z}_3$ operations.
Tuning the effective interpolation between these two gates shows that the $\mathrm{C}_1\mathrm{Z}_3$ gate, although less resourceful~\cite{Chitambar2019}, is physically harder to achieve.
By varying inter-ion interactions, we have shown that optimal gate fidelity is achieved when the next-nearest-neighbor interaction is either negligible or matches the nearest-neighbor one.
Interaction regimes with symmetric couplings are naturally supported by two-dimensional Rydberg-ion crystal geometries.
Recent studies have shown that micromotion-induced effects in such configurations can be quantitatively controlled and do not constitute a fundamental obstacle for coherent Rydberg excitation when operating in appropriate parameter regimes.~\cite{Martins2025}.
Two-dimensional Rydberg-ion crystals therefore represent a promising approach for future exploration.

Moreover, we have demonstrated that the native CCZ gate developed in this work constitutes a useful resource for QEC applications by enabling, for example, a FT implementation of the measurement-free nine-qubit Bacon-Shor code within the available Rydberg-ion gate set.
In this implementation, the CCZ operation is essential for applying the coherent correction.
Simulating the performance of the developed QEC circuit, we observe the quadratic logical-error-rate scaling expected for a FT circuit.
However, to achieve logical error rates below $10^{-6}$, as required for quantum-advantage demonstrations based on logical-qubit algorithms, the physical gate errors would need to be reduced by roughly three orders of magnitude.
Such low error rates are currently beyond experimental reach, and the limited connectivity of a linear ion chain together with the substantial SWAP overhead make a direct extension of this scheme to larger code distances impractical.
These findings suggest that further progress towards experimentally relevant logical error rates will likely require architectures with higher connectivity. 
Two- and three-dimensional Rydberg-ion arrangements naturally provide higher connectivity and therefore are promising to enable larger-distance QEC codes with improved robustness.
In particular, emerging 2D trapped-ion platforms with controllable intra- and inter-chain interactions provide a promising avenue for scaling Rydberg-ion-based QEC~\cite{Jain2024, Valentini2025}. 
In such higher-dimensional settings, the native CCZ gate developed in this work remain a valuable resource, providing efficient multi-qubit interactions directly applicable to larger-scale NISQ algorithmic demonstrations as well as advanced, scalable FT QEC protocols.\\

\section*{Acknowledgments}
We thank D. Locher, M. Mallweger and M. Hennrich for fruitful discussions. 
We gratefully acknowledge the support by the European Union’s Horizon Europe Research and Innovation Program under Grant No. 101046968 (BRISQ) and funding from the Deutsche Forschungsgemeinschaft (DFG, German Research Foundation), and through the Research Unit FOR 5413/1, Grant No. 465199066 and the Federal Ministry of Research, Technology and Space of Germany (BMFTR) through the project NeuQuant. 
KB, TLMG, and MM gratefully acknowledge funding and support provided by the Deutsche Forschungsgemeinschaft (DFG, German Research Foundation) under Germany’s Excellence Strategy ‘Cluster of Excellence Matter and Light for Quantum Computing (ML4Q) EXC 2004/1’ 390534769, the ERC Starting Grant QNets through Grant No. 804247, the Federal Ministry of Research, Technology and Space of Germany (BMFTR) through the project SQale, and the Entangled Logical Qubits (ELQ) program, managed by the Intelligence Advanced Research Projects Activity (IARPA).
This research is also part of the Munich Quantum Valley (K-8), which is supported by the Bavarian state government with funds from the Hightech Agenda Bayern Plus.
The authors gratefully acknowledge the computing time provided to them at the NHR Center NHR4CES at RWTH Aachen University (project number p0020074). 
This work was also supported by funding as part of the Excellence Strategy of the German Federal and State Governments, in close collaboration with the University of T{\" u}bingen and the University of Nottingham. 
This work also received support from the Engineering and Physical Sciences Research Council, Grant No. EP/V031201/1 and EP/W015641/1. JW was supported by the University of T{\" u}bingen through a Research@T{\" u}bingen fellowship. 

\section*{Data and Code Availability}
The code and data used to generate the results presented in this article are available on Zenodo~\cite{Zenodo2025}.

\bibliographystyle{apsrev4-2}
\bibliography{Literature}

\onecolumngrid
\clearpage
\appendix

\section{Gate vs. State Fidelity}\label{app:fidelity}
As mentioned in the main text, we use the state fidelity $\mathcal{F} = |\!\braket{\Psi_{\mathrm{T}}|\Psi(\tau)}\!|^{2}$ for the gate optimization.
In the following we show that the state fidelity between the target state $\ket{\Psi_\mathrm{T}}=U_\mathrm{T}\ket{+}^{\otimes n}$ with $\ket{+}=\frac{1}{\sqrt{2}}(\ket{0}+\ket{1})$ and the system's state $\ket{\Psi(\tau)} = U(\tau)\ket{+}^{\otimes n}$ determined by the unitary time evolution under the Rydberg Hamiltonian~\eqref{eqn:Hamiltonian}, is equivalent to the gate fidelity 
\begin{equation}\label{eqn:gate_fidelity}
    \mathcal{F}_\mathrm{G} = \frac{1}{d}\big|\mathrm{Tr}[U^\dag_\mathrm{T}U(\tau)]\big|
\end{equation}
if the target gate $U_\mathrm{T}$ and the time-evolution operator $U(\tau)$ are both diagonal in the computational basis.
Here, $d=2^n$ denotes the Hilbert-space dimension of an $n$-qubit system.
Inserting $\ket{+}^{\otimes n} = \frac{1}{\sqrt{d}}\sum_{j\in\{0,1\}^{n}}\ket{j}$ into the state fidelity yields
\begin{align}
    |\!\braket{\Psi_\text{T}|\Psi(\tau)}\!|^2 &= \big|\!\bra{+}^{\otimes n}U_\mathrm{T}^\dag U(\tau)\ket{+}^{\otimes n}\!\big|^2 \\
    &= \Big |\frac{1}{d}\sum_{i=0}^{d-1}\sum_{j=0}^{d-1} \braket{i|U_\mathrm{T}^\dag U(\tau)|j}\!\Big |^2\\
    & = \Big |\frac{1}{d}\sum_{i=0}^{d-1}\sum_{j=0}^{d-1}\sum_{k=0}^{d-1} \braket{i|U_\mathrm{T}^\dag \ket{k}\!\!\bra{k}U(\tau)|j}\!\Big |^2, \label{eqn:sum1}
\end{align}
where in the last step we used the completeness of the computational basis $\mathbb{1}=\sum_{k=0}^{d-1}\ket{k}\!\!\bra{k}$.
For the case where the target gate is a CCZ gate, $U_\mathrm{T}$ is diagonal and therefore only matrix elements with $i=k$ contribute to the sum.
Recalling the Hamiltonian~\eqref{eqn:Hamiltonian}, we find that the ground state $\ket{0}$ is isolated in each of the three subsystems. As a consequence the time-evolution $U(\tau)$ must also be diagonal in the computational basis state and all parts of the sum in Eq.~\eqref{eqn:sum1} with $k\neq j$ vanish.
After eliminating two of the sums, we get
\begin{align}
    |\!\braket{\Psi_\text{B}|\Psi(\tau)}\!|^2 &= \Big |\frac{1}{d}\sum_{i=0}^{d-1}\sum_{j=0}^{d-1}\sum_{k=0}^{d-1} \delta_{ik}\delta_{kj}\braket{i|U_\mathrm{T}^\dag \ket{k}\!\!\bra{k}U(\tau)|j}\Big |^2 \\
    &= \Big |\frac{1}{d}\sum_{i=0}^{d-1}\braket{i|U_\mathrm{T}^\dag \ket{i}\!\!\bra{i}U(\tau)|i}\Big |^2 \\
    &= \Big |\frac{1}{d}\mathrm{Tr}[U_\mathrm{T}^\dag U(\tau)]\Big |^2,
\end{align}
which corresponds to the squared gate fidelity~\eqref{eqn:gate_fidelity}.
Even though the gate and state fidelity are not equal, but differ by a power of two, they are equivalent in the sense that they contain the same information and serve as valid cost functions in the gate pulse profile optimization process.

\section{Independence of Phase and Population Error}\label{app:errors}
In the main text, we have defined the phase and population error and discussed them independently from each other.
The following calculation shows that infidelity caused by population or phase incongruence is indeed independent of the other in second order.
The state fidelity can be written in terms of its real and imaginary part
\begin{equation}\label{eqn:state_fidelity_app}
    \mathcal{F} = |\!\braket{\Psi_\mathrm{T}|\Psi(\tau)}\!|^2 = \Re(\braket{\Psi_\mathrm{T}|\Psi(\tau)})^2 + \Im(\braket{\Psi_\mathrm{T}|\Psi(\tau)})^2,  
\end{equation}
where $\Psi_\mathrm{T}=\sum_{abc}c_{abc}^\mathrm{opt}e^{i\phi_{abc}^\mathrm{opt}}\ket{abc},\, c_{abc}^\mathrm{opt}, \phi_{abc}^\mathrm{opt}\in \mathbb{R}$ is the target state with $a, b, c \in \{0,1\}$ and $\Psi(\tau)=\sum_{abc}c_{abc}e^{i\phi_{abc}}\ket{abc},\, c_{abc}, \phi_{abc}\in \mathbb{R}$ is the system’s state after evolving for a time  $\tau$ under the considered Hamiltonian which is intended to realize the desired gate $\Psi_\text{T} = U_\text{opt}\ket{\Psi(0)}$.
We assume that the population and phases deviate from their optimal value by a small error 
\begin{align}
\phi_{abc} &= \phi_{abc}^\mathrm{opt}+\delta\phi_{abc},\quad |\delta\phi_{abc}|\ll 1, \\
c_{abc} &= c_{abc}^\mathrm{opt}-\delta c_{abc},\quad 0<\delta c_{abc}\ll1.
\end{align}
With that we calculate the overlap between the two states in Eq.~\eqref{eqn:state_fidelity_app} up to second order in the errors
\begin{align}
\braket{\Psi_\mathrm{T}|\Psi(\tau)} &= \sum_{abc}c_{abc}^\mathrm{opt}c_{abc}e^{i\phi_{abc}}e^{-i\phi_{abc}^\mathrm{opt}}\\
&= \sum_{abc}c_{abc}^\mathrm{opt}(c_{abc}^\mathrm{opt}-\delta c_{abc})e^{i\delta\phi_{abc}}\\
& =\sum_{abc}c_{abc}^\mathrm{opt}(c_{abc}^\mathrm{opt}-\delta c_{abc})\big(1 + i\delta\phi_{abc}-\frac{\delta\phi_{abc}^2}{2} + \mathcal{O}(\delta^3)\big)\\
&=1-\sum_{abc}\big(c_{abc}^\mathrm{opt}\delta c_{abc}+(c_{abc}^\mathrm{opt}\big)^2\frac{\delta\phi_{abc}^2}{2}) + i \sum_{abc}\big((c_{abc}^\mathrm{opt})^2-c_{abc}^\mathrm{opt}\delta c_{abc}\big)\delta\phi_{abc}\big)+ \mathcal{O}(\delta^3), \label{eqn:overlap}
\end{align}
where in the second step we expanded the exponential function and in the last step we used the normalization of the target state $\sum_{abc}(c_\mathrm{abc}^\mathrm{opt})^2 =1$. Note that $\mathcal{O}(\delta^n)$ denotes the $n$th order of the error independent whether phase or population error, e.g. $\delta c_{abc}\delta \phi_{abc}^2 \sim\mathcal{O}(\delta^3)$. 
From Eq.~\eqref{eqn:overlap}, we read off the real and imaginary parts, insert these into the fidelity and again expand up to second order
\begin{align}
F =& \Big(1-\sum_{abc}\big(c_{abc}^\mathrm{opt}\delta c_{abc}+(c_{abc}^\mathrm{opt}\big)^2\frac{\delta\phi_{abc}^2}{2})+ \mathcal{O}(\delta^3)\big)\Big)^2
+\Big( \sum_{abc}\big((c_{abc}^\mathrm{opt})^2-c_{abc}^\mathrm{opt}\delta c_{abc}\big)\delta\phi_{abc} + \mathcal{O}(\delta^3)\big)\Big)^2\\
=&\Big(1-\sum_{abc}c_{abc}^\mathrm{opt}\delta c_{abc}\Big)^2-\sum_{abc}(c_{abc}^\mathrm{opt})^2\delta \phi_{abc}^2 + \Big(\sum_{abc}(c_{abc}^\mathrm{opt})^2\delta\phi_{abc}\Big)^2 + \mathcal{O}(\delta^3).
\end{align}
We see that in second order the phase and population errors are independent from each other. 
Furthermore, we find that the phase error $\delta\phi_\text{abc}$ does not contribute in first order.

\section{Optimization Parameters}\label{app:parameter}
All gate optimizations were performed within the parameter ranges
\begin{equation}
    \Omega_0 \in [0,4]\,V, \qquad\delta_0 \in [0,10]\,V, \qquad\Delta_0 \in [-10, 10]\,V.
\end{equation}
Table~\ref{tab:parameter_table} lists the fixed parameters used in the optimizations, the resulting optimized parameters, and the performance parameters underlying the plots in Fig.~\ref{fig:gates} and Fig.~\ref{fig:decay}.

\begin{table}[h]
\centering
\begin{tabular}{ll|cccc}
\textbf{Protocol}                 &  & \hspace{0.3cm}\textbf{Fig.~\ref{fig:gates}(b)}\hspace{0.3cm} & \hspace{0.3cm} \textbf{Fig.~\ref{fig:gates}(c)}\hspace{0.3cm} & \hspace{0.3cm}\textbf{Fig.~\ref{fig:decay}(a)}\hspace{0.3cm} & \hspace{0.3cm} \textbf{Fig.~\ref{fig:decay}(b)}\hspace{0.3cm}     \\ \hline \hline
                                  &                                    &                       &                         &                                      &                                       \\[-0.2cm]
    \textbf{Fixed}                & $\tau\,[2\pi/V]$                   & $87.5$                & $25.0$                  & $87.5$                               & $87.5$                                \\
    \textbf{parameters}           & $\Omega_\mathrm{MW}\,[V]$          & $20.0$                & $20.0$                  & $20.0$                               & $20.0$                                \\
                                  & $\gamma_\mathrm{R}\,[V/2\pi]$      & $0.0$                 & $0.0$                   & $1.64\cdot10^{-3}$                   & $1.64\cdot10^{-3}$                    \\
                                  &                                    &                       &                         &                                      &                                       \\[-0.2cm]
    \textbf{Gate}                 & $\Omega_0^{\mathrm{opt}}\,[V]$     & $3.55$                & $1.19$                  & $3.55$                               & $0.53$                                \\
    \textbf{parameters}           & $\delta_0^{\mathrm{opt}}\,[V]$     & $8.34$                & $6.81$                  & $8.34$                               & $5.76$                                \\
                                  & $\Delta_0^{\mathrm{opt}}\,[V]$     & $-4.09$               & $-3.25$                 & $-4.09$                              & $-4.05$                               \\
                                  &                                    &                       &                         &                                      &                                       \\[-0.2cm]
    \textbf{Performance}          & $\bar{p}$                          & $1.09\cdot10^{-8}$    & $6.33\cdot 10^{-5}$     & $1.03\cdot 10^{-1}$                  & $1.69\cdot 10^{-2}$                   \\
                                  & $\bar{\varphi}$                    & $9.48\cdot 10 ^{-7}$  & $3.24\cdot10^{-2}$      & $9.48\cdot 10^{-7}$                  & $1.03\cdot 10^{-2}$                   \\
                                  & $\mathcal{F}$                      & $1-9.6\cdot10^{-7}$   & $96.75\,\%$             & $89.65\,\%$                          & $97.29\,\%$                               \\
\end{tabular}
\caption{Overview of fixed parameters used in the optimization, resulting optimized parameters, and performance metrics associated with Fig.~\ref{fig:gates} and Fig.~\ref{fig:decay}.}
\label{tab:parameter_table}
\end{table}

\section{\texorpdfstring{Implementation of a Native $\mathrm{C}_1\mathrm{Z}_3$ Gate}{Implementation of a native C₁Z₃ gate}}\label{app:C1Z3}

\begin{figure}[t]
    \centering
    \includegraphics[scale = 1]{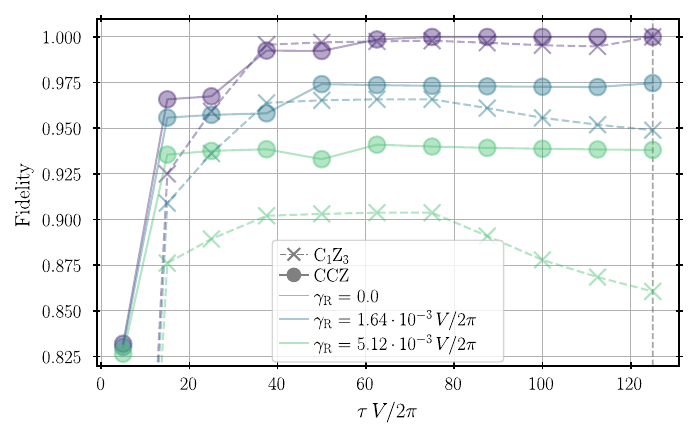}
 \caption{$\mathrm{C}_1\mathrm{Z}_3$-gate optimization results. 
 Comparison of the fidelity of optimized $\mathrm{C}_1\mathrm{Z}_3$- and CCZ-gate protocols~\eqref{eqn:pulses} vs. gate duration $\tau$ for different decay rates $\gamma_\mathrm{R}$.
 The grey vertical dashed line indicates the three gates plotted in Fig.~\ref{fig:C1Z3_gates}.}
  \label{fig:C1Z3_fidelity}
\end{figure}
\begin{figure}[t]
    \centering
    \includegraphics[scale = 1]{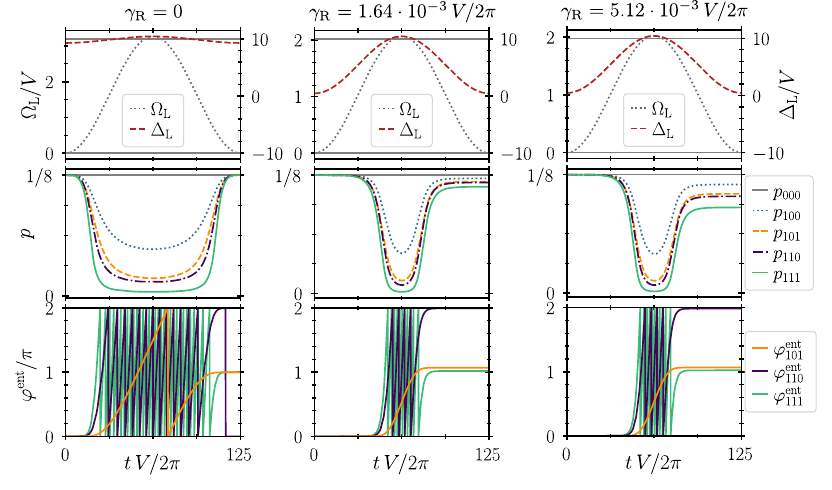}
 \caption{Gate dynamics for three optimized $\mathrm{C}_1\mathrm{Z}_3$ gates for different decay rates $\gamma_\mathrm{R}$ with fidelities $F\approx 100\,\%,\, 94.89\,\%$ and $86.06\,\%$, respectively.
  Top: Modulation of the Rabi frequency $\Omega_\mathrm{L}$ and the detuning $\Delta_\mathrm{L}$ of the excitation laser. 
  The gray lines at $\Delta_\mathrm{L}=\mp\frac{\Omega_\mathrm{MW}}{2}$ indicate resonance with the dressed Rydberg states $\ket{\mathcal{D}_\pm}$.
  Center: Population dynamics of the computational basis states for the initial state $\ket{+++}$.
  Bottom: Accumulated entangling phases $\varphi_{abc}^\mathrm{ent}$ with $(a,b,c)\in\{0,1\}^3$.}
  \label{fig:C1Z3_gates}
\end{figure}

In the following we briefly discuss the implementation of a $\mathrm{C}_1\mathrm{Z}_3$ gate in our setup.
Figure \ref{fig:C1Z3_fidelity} depicts the fidelities of the resulting $\mathrm{C}_1\mathrm{Z}_3$-gate implementations as well as of the CCZ gates as given in Fig.~\ref{fig:gates}.
While the $\mathrm{C}_1\mathrm{Z}_3$ gate exhibits only marginally reduced performance in the absence of decay, its fidelities are substantially degraded compared to the CCZ gate when decay is present.
For a decay rate of $\gamma_\mathrm{R} = 1.64 \cdot 10^{-3}\,V/2\pi$ ($\gamma_\mathrm{R} = 5.12 \cdot 10^{-3}\,V/2\pi$), the best fidelity achieved for a $\mathrm{C}_1\mathrm{Z}_3$ gate is $96.58\,\%$ ($90.38\,\%$), which is below the best fidelity of the CCZ gates of $97.47\,\%$ ($94.10\,\%$).
The reason why the $\mathrm{C}_1\mathrm{Z}_3$ gate performs worse especially for higher decay rates, can be understood by having a closer look at the gates as depicted in Fig.~\ref{fig:C1Z3_gates}.
In the case with no decay and whenever the gate time is long enough the ions can spend a sufficient amount of time in the Rydberg states to accumulate the desired phases and reach close-to-unit fidelity similar to the case in Fig~\ref{fig:gates}(b).
Whenever decay is present, we observe that the time in Rydberg states is reduced to minimize population errors, yet remains sufficiently long to accumulate the required phase $\varphi_{101}^\mathrm{ent}\approx \pi$. 
In contrast to the CCZ gate, keeping $\varphi_{101}^\mathrm{ent}$ as small as possible is not a good strategy for the $\mathrm{C}_1\mathrm{Z}_3$ gate, since doing so would cause large phase errors. 
The ions must therefore spend adequate time in Rydberg states, resulting in population errors that significantly exceed those of the CCZ gate.
Since the population error grows with increasing decay rate, the performance is especially poor at $\gamma_\mathrm{R} = 5.12 \cdot 10^{-3}\,V/2\pi$.
Detailed investigations of the $\mathrm{C}_1\mathrm{Z}_3$ implementation with trapped Rydberg ions and exploration of improvement strategies are reserved for future work.

\section{CCZ-Gate Decompositions}\label{app:decom}
\begin{figure}[t]
    \centering
    \includegraphics[scale = 0.8]{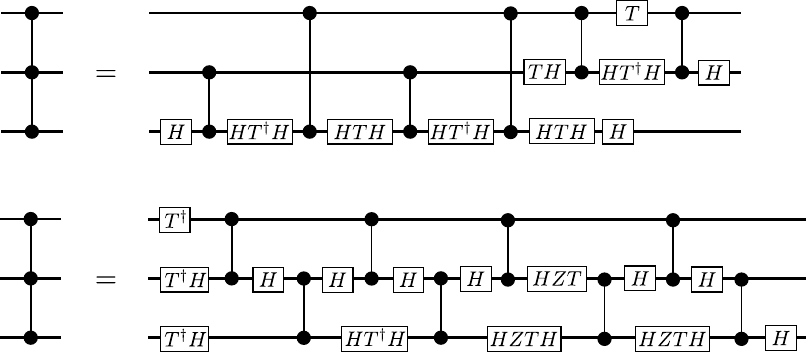}
    \caption{Two possible decompositions of the CCZ gate.  
    The upper circuit shows the standard textbook decomposition~\cite{Shende2009} consisting of six CZ gates, four between nearest neighbours and two between next-nearest neighbours, together with single-qubit rotations.  
    The lower circuit implements the CCZ using only nearest-neighbour CZ gates and single-qubit rotations~\cite{Gwinner2021}.}
  \label{fig:decomposition}
\end{figure}

In the following, we introduce the CCZ-gate decomposition as used for the benchmarking in the main text~\cite{Shende2009, Gwinner2021, Nakanishi2024, Sun2024}.  
Figure~\ref{fig:decomposition} illustrates two possible ways to decompose the CCZ gate.  
The upper circuit shows a standard textbook approach, which employs six CZ gates, four between nearest neighbours and two between next-nearest neighbours, together with several single-qubit rotations.  
The lower circuit implements the same operation using only nearest-neighbour CZ gates, eight in total, along with single-qubit rotations.  
Since next-nearest-neighbour CZ gates generally exhibit lower fidelities than nearest-neighbour gates, we use the latter approach for benchmarking in this work.
Further, we assume one layer of single-qubit gates with an execution time of $1\,\text{\textmu}\mathrm{s}$ between the two-qubit gates.
In practice, the number and type of single-qubit gates depend on the available native gate set.
Therefore, this approximation yields a lower bound on the total execution time, which is sufficient for the benchmarking discussed in the main text. 

\section{Measurement-Free FT Implementation of the Bacon-Shor QEC Circuit on a Rydberg-Ion Platform}\label{app:BS-circuit}
\begin{figure}[t]
    \centering
    \includegraphics[scale = 1]{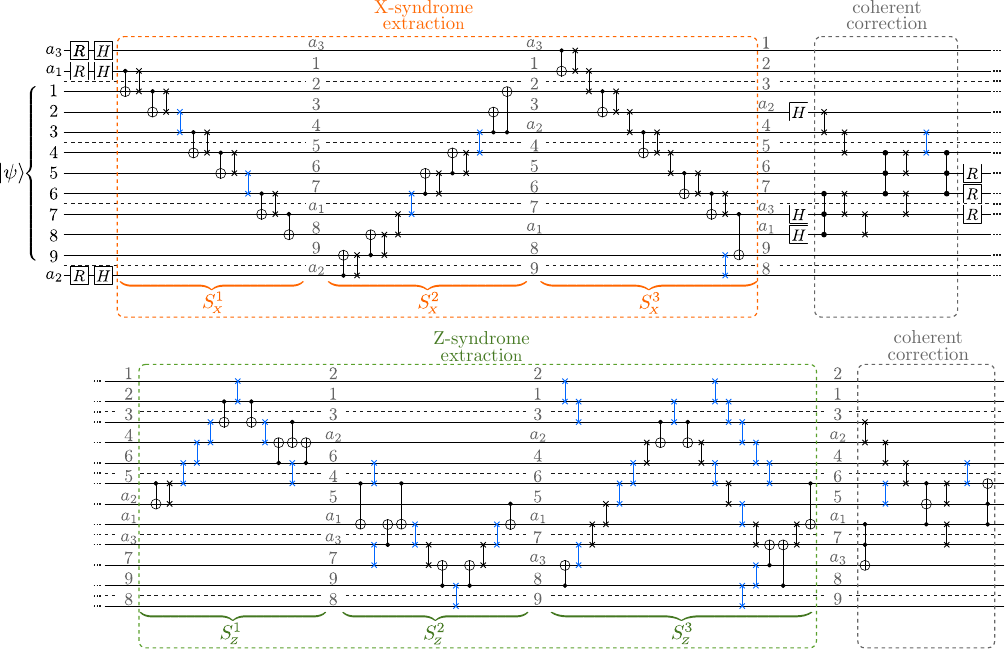}
    \caption{Measurement-free FT QEC circuit for the nine-qubit Bacon-Shor code implemented with the Rydberg-ion gate set.  
    In the first block, the $X$-type stabilizers are extracted, followed by the corresponding coherent $Z$-correction and reset of the ancilla qubits.  
    The $Z$-type stabilizers are then coherently mapped to the ancilla qubits, after which the final $X$-correction is applied.  
    Fault-tolerant SWAP gates, indicated by blue SWAP symbols, are used throughout the circuit, where needed, to ensure circuit-level fault tolerance.  
    The black dashed lines represent the ancillary qubits needed for the FT SWAP operation.
    Grey numbers indicate qubit positions along the ion string and are included to help track the qubits as they are permuted throughout the circuit.  
    All multi-qubit gates correspond to the native Rydberg-ion gate set $\{\mathrm{CZ}, \mathrm{CCZ}\}$ up to local Hadamard rotations; CNOT decompositions are omitted for clarity.  
    Note that a CNOT followed by a non-FT SWAP can be reduced to two CNOT operations, however, the SWAP gates are kept explicitly for better readability. 
    While the circuit is not fully optimized, the number of required FT SWAPs has been minimized while preserving overall fault tolerance. 
    Compared to a naive implementation employing FT SWAPs throughout, this optimization reduces the total number of physical CNOT gates by approximately $25\,\%$.}
  \label{fig:full_BS}
\end{figure}

Figure~\ref{fig:full_BS} shows the complete circuit of the FT implementation of the Bacon-Shor code introduced in Fig.~\ref{fig:FT_BSC}, expressed entirely in terms of the native Rydberg-ion gate set.  
In the first stage, the $X$-type stabilizers are coherently extracted, followed by the coherent correction of potential $Z$-errors and a reset of the ancilla qubits. 
The second stage performs the corresponding $Z$-type stabilizer extraction and the final $X$-error correction.
Fault-tolerant SWAP gates are inserted throughout the circuit to ensure circuit-level fault tolerance within the limited connectivity of a linear ion string.

\begin{figure}[t]
    \centering
    \includegraphics[scale = 1]{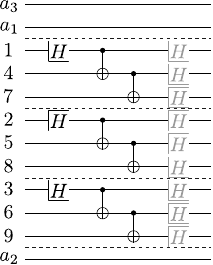}
    \caption{Fault-tolerant circuit to encode the logical states of the Bacon-Shor code \cite{Egan2021}.   
    The logical state $\ket{0}_\mathrm{L}$ ($\ket{+}_\mathrm{L}$) is prepared by generating GHZ-like states along each row (column) of the data-qubit lattice.  
    For encoding $\ket{+}_\mathrm{L}$, the transversal Hadamard gate (grey) is omitted.  
    Errors occurring during the encoding can propagate only within a single GHZ state and therefore never lead to a logical error, implying that the encoding circuit is FT.}
  \label{fig:encoding}
\end{figure}

Figure~\ref{fig:encoding} shows the FT circuit used to prepare the logical states of the nine-qubit Bacon-Shor code.  
The state~$\ket{0}_\mathrm{L}$ ($\ket{+}_\mathrm{L} = \frac{1}{\sqrt{2}}[\ket{0}_\mathrm{L}+\ket{1}_\mathrm{L}]$) is generated by creating GHZ-like states along the rows (columns) of the data-qubit lattice, with the transversal Hadamard gate included only for the $\ket{0}_\mathrm{L}$ encoding.  
Since mid-circuit errors can propagate only within a single GHZ state, no logical errors can result, and the encoding process is FT.

\end{document}